\documentclass[11pt]{article}

\usepackage{amssymb,amsthm,amsmath}
\usepackage[mathscr]{eucal}
\usepackage[round]{natbib}

\numberwithin{equation}{section} 

\usepackage{bm} 

\newcommand{\R}{\mbox{$\mathbb R$}}
\newcommand{\Z}{\mbox{$\mathbb Z$}}

\newcommand{\sA}{\mbox{$\mathscr A$}}
\newcommand{\sC}{\mbox{$\mathscr C$}}
\newcommand{\sD}{\mbox{$\mathscr D$}}

\newcommand{\ssA}{\scriptstyle{\mathscr A}}
\newcommand{\ssC}{\scriptstyle{\mathscr C}}
\newcommand{\ssD}{\scriptstyle{\mathscr D}}

\newcommand{\cart}[3]{\ensuremath{{#1}_{#2} \times \cdots \times {#1}_{#3}}}
\newcommand{\veeA}[1]{\ensuremath{\vee_{i \in {#1}}\sA_i}}
\newcommand{\pperp}{\ensuremath{{\rm \perp}\kern-.60em {\rm \perp} }}
\newcommand{\npperp}{\ensuremath{{\rm \perp}\kern-.60em {\rm \not\perp} }}
\newcommand\set[1]{\{1, \ldots, #1 \}}
\newcommand\norm[1]{\| #1 \|}

\newcommand{\Uhat}{\ensuremath{\hat{U}}}
\newcommand{\uhat}{\ensuremath{\hat{u}}}

\newtheorem{theorem}{Theorem}[section]
\newtheorem{lemma}[theorem]{Lemma}
\newtheorem{corollary}[theorem]{Corollary}
\newtheorem{definition}[theorem]{Definition}
\newtheorem{example}[theorem]{Example}
\newtheorem{remark}[theorem]{Remark}

\begin{document}

\author{Jay H. Beder
\\Department of Mathematical Sciences\\University of Wisconsin-Milwaukee\\
P.O. Box 413\\Milwaukee, WI 53201-0413\\
{\tt beder@uwm.edu}\\ \ \\Wiebke S. Diestelkamp\footnote{Funded in
part by an NSF-AWM Mentoring Travel Grant and a University of Dayton
Women's Center Research Grant.}
\\Department of Mathematics\\University of Dayton\\
Dayton, OH 45469-2316\\
{\tt wiebke@udayton.edu}\\ \ \\ \ }
\title{Box-Hunter resolution in nonregular fractional factorial designs}
\date{}

\maketitle

\vspace{-1cm}

\begin{abstract}
In a 1961 paper, Box and Hunter defined the resolution of a regular fractional
factorial design as a measure of the amount of aliasing in the fraction. They
also indicated that the maximum resolution is equal to the minimum length of a
defining word.  The idea of a wordlength pattern has now been extended to
nonregular designs by various authors, who show that the minimum generalized
wordlength equals the maximum strength plus 1.

Minimum generalized wordlength is often taken as the definition of resolution. However, Box and Hunter's original definition, which does not depend on wordlength, can be extended to nonregular designs if they are simple. The purpose of this paper is to prove that the maximum Box-Hunter resolution does equal the maximum strength plus 1, and therefore equals the minimum generalized wordlength. Other approaches to resolution are briefly discussed.

\end{abstract}


{\footnotesize {\bf AMS(MOS) subject classification.}
Primary: 62K15; 
Secondary:
05B15, 
62K05 
}
\section{Introduction}\label{intro}

\citet{BoxHunt61}  introduced the notion of resolution of a regular fraction,
and observed that the maximum resolution, say $R_{\max}$, is equal to the
length of the shortest defining word.  \citet{FriesHunter80} pointed out that
the \emph{number} of defining words of length $R_{\max}$ discriminates between
fractions of equal resolution in a useful way: for example, for regular
fractions of equal size and having the same factors, fewer defining words of
length $R_{\max}$ implies less aliasing between main effects and interactions
of order $R_{\max}-1$.  This led them to consider the \emph{wordlength pattern}
$(A_1, \ldots, A_k)$ of a regular fraction having $k$ factors, where $A_i$ is
the number of defining words of length $i$, and to introduce their criterion of
\emph{relative aberration} for comparing two designs.


There have been various proposals for extending the idea of wordlength patterns
to non-regular (and possibly mixed-level) designs. \citet{DengTang99} and
\citet{TangDeng99} gave a definition for 2-level designs that was generalized
to arbitrary mixed-level designs by \citet{XuWu01}.  An equivalent
coding-theoretic version was developed simultaneously by
\citet{MaFang01}.  All these authors prove that the minimum generalized
wordlength satisfies
  \begin{equation} \label{mini=tmax} \min \{i: A_i > 0 \} = t_{\max} + 1, \end{equation}
$t_{\max}$ being the maximum \emph{strength} of the design (considered as an
orthogonal array).

With some variation, these authors\footnote{See \citet[page 88]{MaFang01} and \citet[page 1068]{XuWu01}, as well as \citet[Definition 4.1]{ChengYe04}.  A variation by Deng and Tang is noted in Remark~\ref{strength-rem} below.} simply \emph{define} resolution to be the left-hand side of (\ref{mini=tmax}).  However, Box and Hunter have given us an independent
concept of resolution:
  \begin{quote} A design has resolution $R$ if every interaction of $p$ factors
  is unaliased with every interaction of fewer than $R-p$ factors.\end{quote}
This definition, and the definition of aliasing that underlies it, have been
extended to arbitrary simple (or equireplicate) designs in a previous paper
\citep{bederrao2}, where it was shown that a simple design of strength $t$ has
``Box-Hunter" resolution $R \ge t+1$ (Corollary~\ref{coro} below). In this
paper we show that $R_{\max} = t_{\max}+1$ (Theorem~\ref{strength/resolution}).
This and (\ref{mini=tmax}) then \emph{prove} that $R_{\max} =\min \{i: A_i > 0
\}$.

To be sure, this result depends on the particular definitions of aliasing and
resolution that we are using. Other approaches to these concepts will be
discussed in Section~\ref{approaches}. The present approach is illustrated with
a regular fraction in Section~\ref{example} in preparation for the abstract
set-up in Section~\ref{resolu}.


\medskip
%
%

\textbf{Notation and basic definitions.}  We follow the notation and
definitions given in \citet{bederrao2}.  In particular, the cardinality of a
set $E$ is denoted by $|E|$, and the empty set by $\emptyset$.  The integers
are denoted by \Z, and the integers modulo $n$ by $\Z/n$.  The real numbers are
denoted by \R, and the real-valued functions on the set $T$ by $\R^T$. Given
any finite set $T$ (for us, the set of treatment combinations), $\R^T$ is a
Euclidean space with inner product
  \begin{equation} \label{inner-eq} (u,v) = \sum_{s \in T} u(s) v(s)
  \end{equation}
for $u, v \in \R^T$ and norm $\norm{v} = \sqrt{(v,v)}$.  If we fix an
ordering of the elements of $T$, we may view $u$ and $v$ as ordinary
column vectors in the Euclidean space $\R^g$, where $g=|T|$. Then the
formula in (\ref{inner-eq}) is the ordinary dot product.

We denote by 1 the constant function taking the value 1, and by $1_C$
the indicator or characteristic function of the set $C \subset T$:
\[ 1_{C}(s) = \left\{ \begin{array}{ll} 1 & \mbox{if $s \in C$}, \\ 0 &
\mbox{if $s \not\in C$}.  \end{array} \right. \] Thus 1 is $1_T$.
Note that $(1_C, 1_D) = |C \cap D|$.

If there are $k$ factors whose levels  are indexed by  sets $A_1, \ldots, A_k$
of size $s_1, \ldots, s_k$, respectively,  then the set of treatment
combinations (or \emph{cells}) is $T = \cart{A}{1}{k}$. We will refer to $T$ as
the \emph{full factorial design}.  A \emph{fractional factorial design}, or
\emph{fraction}, is one in which each treatment combination appears with some
multiplicity (possibly 0).  The design is \emph{simple} if it is a subset $S$
of $T$, that is, if each treatment combination used in the design appears only
once.

The design $T$ is \emph{symmetric} if  $s_1 = \cdots = s_k = s$, in
which case we may take $A_1 = \cdots = A_k = A$;  otherwise it is
\emph{asymmetric} or \emph{mixed-level}.  Similar terminology applies
to a fraction.  If in a symmetric design $s$ is a prime or prime
power, we may take $A$ to be the finite field $GF(s)$. In this case
the fraction is \emph{regular} if it is the solution set of a system
of linear equations over the finite field $GF(s)$.

If the cells of the fraction are written as rows or columns of a matrix, then
the fraction is an orthogonal array and thus has strength $t$, for some $t$,
and (in the symmetric case) index $\lambda$ (see Section~\ref{resolu}).

Other notation is introduced as needed.

\section{An illustrative example} \label{example}
In this section we illustrate the abstract definition of aliasing that will
follow in Section~\ref{resolu}.  This is often illustrated in introductory
experimental design texts by a regular $2^{3-1}_{\rm III}$ fraction.  A regular
$3^{3-1}_{\rm III}$ fraction will better display the features of the general
situation.

Consider a regular fraction with defining relations
%
\[ \begin{array}{cll} I &= AB^2C^2 &(= A^2 BC) \end{array} \]
and aliases
\[ \begin{array}{cll}
A &= ABC &= BC \\
B &= AC^2 &= ABC^2 \\
C &= AB^2 &= AB^2C \\
AB &= AC &= BC^2.
\end{array} \]
There are three such fractions,  each having (maximum) resolution 3. We will
choose the one given as the solution set to $x + 2y + 2z \equiv 1 \pmod{3}$,
namely the cells
  \[ 002, \; 011, \; 020, \; 100, \; 112, \; 121, \; 201, \; 210, \; \mbox{and} \; 222. \]

To construct it, we create a pair of contrasts for each of the 13 main effects
and components of interaction in the full $3^3$ factorial.  Each effect is
described by a partition of the 27 treatment combinations into 3 blocks, and we
create a pair of contrasts by assigning $1, -1, 0$ and $1, 0 -1$, respectively,
to the blocks.  A portion of the resulting contrast vectors (of length 27)
would look like this: \bigskip

\setlength{\tabcolsep}{1mm}
\begin{tabular}{c|rr|rr|rr|rr|rr|rr|rr|rr|rr|rr|rr|rr|rr}
  cell  &\multicolumn{2}{c}{$A$}&\multicolumn{2}{c}{$B$}&\multicolumn{2}{c}{$C$}&
  \multicolumn{2}{c}{$AB$}&\multicolumn{2}{c}{$AB^2$}&\multicolumn{2}{c}{$AC$}&\multicolumn{2}{c}{$AC^2$} &\multicolumn{2}{c}{$BC$} & \multicolumn{2}{c}{$BC^2$} &     \multicolumn{2}{c}{$ABC$} &
  \multicolumn{2}{c}{$ABC^2$} &\multicolumn{2}{c}{$AB^2C$} &\multicolumn{2}{c}{$AB^2C^2$}
\\
 000 & 1 & 1 & 1 & 1 & 1 & 1 & 1 & 1 & 1 & 1 & 1 & 1 & 1 & 1 & 1 & 1 & 1
   & 1 & 1 & 1 & 1 & 1 & 1 &  1 & 1 & 1  \\
 001 & 1 & 1 & 1 & 1 & -1 & 0 & 1 & 1 & 1 & 1 & -1 & 0 & 0 & -1 & -1 & 0 & 0
  & -1 & -1 & 0 & 0 & -1 & -1 &  0 & 0 & -1  \\
 002 & 1 & 1 & 1 & 1 & 0 & -1 & 1 & 1 & 1 & 1 & 0 & -1 & -1 & 0 & 0 & -1 & -1
   & 0 & 0 & -1 & -1 & 0 & 0 & -1 & -1 & 0  \\
 010 & 1 & 1 & -1 & 0 & 1 & 1 & -1 & 0 & 0 & -1 & 1 & 1 & 1 & 1 & -1 & 0 & -1
  & 0 & -1 & 0 & -1 & 0 & 0 & -1 & 0 & -1  \\
 011 & 1 & 1 & -1 & 0 & -1 & 0 & -1 & 0 & 0 & -1 & -1 & 0 & 0 & -1 & 0 & -1 &1
   & 1 & 0 & -1 & 1 & 1 & 1 &  1 & -1 & 0  \\
 012 & 1 & 1 & -1 & 0 & 0 & -1 & -1 & 0 & 0 & -1 & 0 & -1 & -1 & 0 & 1 & 1 & 0
   & -1 & 1 & 1 & 0 & -1 & -1 &  0 & 1 & 1  \\
 020 & 1 & 1 & 0 & -1 & 1 & 1 & 0 & -1 & -1 & 0 & 1 & 1 & 1 & 1 & 0 & -1 & 0
   & -1 & 0 & -1 & 0 & -1 & -1 &  0 & -1 & 0  \\
 021 & 1 & 1 & 0 & -1 & -1 & 0 & 0 & -1 & -1 & 0 & -1 & 0 & 0 & -1 & 1 & 1 &-1
   & 0 & 1 & 1 & -1 & 0 & 0 & -1 & 1 & 1  \\
 022 & 1 & 1 & 0 & -1 & 0 & -1 & 0 & -1 & -1 & 0 & 0 & -1 & -1 & 0 & -1 & 0 &1
   & 1 & -1 & 0 & 1 & 1 & 1 &  1 & 0 & -1  \\
 100 & -1 & 0 & 1 & 1 & 1 & 1 & -1 & 0 & -1 & 0 & -1 & 0 & -1 & 0 & 1 & 1 & 1
   & 1 & -1 & 0 & -1 & 0 & -1 &  0 & -1 & 0  \\
 101 & -1 & 0 & 1 & 1 & -1 & 0 & -1 & 0 & -1 & 0 & 0 & -1 & 1 & 1 & -1 & 0 & 0
   & -1 & 0 & -1 & 1 & 1 & 0 & -1 & 1 & 1  \\
 102 & -1 & 0 & 1 & 1 & 0 & -1 & -1 & 0 & -1 & 0 & 1 & 1 & 0 & -1 & 0 & -1 &-1
   & 0 & 1 & 1 & 0 & -1 & 1 &  1 & 0 & -1  \\
  \vdots & \vdots & \vdots & \vdots & \vdots & \vdots & \vdots & \vdots &
  \vdots & \vdots & \vdots & \vdots & \vdots & \vdots & \vdots & \vdots & \vdots & \vdots & \vdots & \vdots & \vdots &  \vdots & \vdots & \vdots & \vdots & \vdots & \vdots
\end{tabular}

\bigskip

We now select from these only the 9 treatment combinations in our fraction,
restricting the original columns to those 9 cells.  This yields the columns
below.  Of course, those for $AB^2C^2$ no longer are contrast vectors as they
represent a defining word. \bigskip

\setlength{\tabcolsep}{1mm}
\begin{tabular}{c|rr|rr|rr|rr|rr|rr|rr|rr|rr|rr|rr|rr|rr}
  cell  &\multicolumn{2}{c}{$A$}&\multicolumn{2}{c}{$B$}&\multicolumn{2}{c}{$C$}&
  \multicolumn{2}{c}{$AB$}&\multicolumn{2}{c}{$AB^2$}&\multicolumn{2}{c}{$AC$}&\multicolumn{2}{c}{$AC^2$} &\multicolumn{2}{c}{$BC$} & \multicolumn{2}{c}{$BC^2$} &     \multicolumn{2}{c}{$ABC$} &
  \multicolumn{2}{c}{$ABC^2$} &\multicolumn{2}{c}{$AB^2C$} &\multicolumn{2}{c}{$AB^2C^2$}
\\
  002   &   1  &   1  &   1  &   1  &   0  &  -1  &  1  &  1  &  1 &  1 &  0 & -1 & -1 &  0 &  0 & -1 & -1 &  0 &  0 & -1 & -1 &  0 &  0 & -1 & -1 & 0\\
  011   &   1  &   1  &  -1  &   0  &  -1  &   0  & -1  &  0  &  0 & -1 & -1 &  0 &  0 & -1 &  0 & -1 &  1 &  1 &  0 & -1 &  1 &  1 &  1 &  1 & -1 & 0\\
  020   &   1  &   1  &   0  &  -1  &   1  &   1  &  0  & -1  & -1 &  0 &  1 &  1 &  1 &  1  & 0 & -1 &  0 & -1 &  0 & -1 &  0 & -1 & -1 &  0 & -1 & 0 \\
  100   &  -1  &   0  &   1  &   1  &   1  &   1  & -1  &  0  & -1 &  0 & -1 &  0 & -1 &  0  & 1 &  1 &  1 &  1 & -1 &  0 & -1 &  0 & -1 &  0 & -1 & 0\\
  112   &  -1  &   0  &  -1  &   0  &   0  &  -1  &  0  & -1  &  1 &  1 &  1 &  1 &  0 & -1  & 1 &  1 &  0 & -1 & -1 &  0 &  1 &  1 &  0 & -1 & -1 & 0\\
  121   &  -1  &   0  &   0  &  -1  &  -1  &   0  &  1  &  1  &  0 & -1 &  0 & -1 &  1 &  1  & 1 &  1 & -1 &  0 & -1 &  0 &  0 & -1 &  1 &  1 & -1 & 0\\
  201   &   0  &  -1  &   1  &   1  &  -1  &   0  &  0  & -1  &  0 & -1 &  1 &  1 & -1 &  0  &  -1 &  0 &  0 & -1 &  1 &  1 & -1 &  0 &  1 &  1 & -1 & 0\\
  210   &   0  &  -1  &  -1  &   0  &   1  &   1  &  1  &  1  & -1 &  0 &  0 & -1 &  0 & -1  & -1 &  0 & -1 &  0 &  1 &  1 &  1 &  1 & -1 &  0 & -1 & 0\\
  222   &   0  &  -1  &   0  &  -1  &   0  &  -1  & -1  &  0  &  1 &  1 & -1 &  0 &  1 &  1 &  -1 &  0 &  1 &  1 &  1 &  1 &  0 & -1 &  0 & -1 & -1 & 0
\end{tabular}
\bigskip

The aliasing of $AB$ and $AC$, for example, means that the restricted contrast
vectors for $AC$ are linear combinations of those for $AB$, and vice versa. On
the other hand, the fact that, say, $C$ and $AB$ are \emph{not} aliased in the
fraction means that the restricted contrast vectors for $C$ are
\emph{orthogonal} to those of $AB$. Another way to say this is that the
\emph{span} (in $\R^9$) of the restricted columns for $AB$ equals the span of
those for $AC$, and is orthogonal to the span of those representing $C$.

Of course, the particular choice of contrast vectors representing each effect
in the full factorial is not the essential thing.  Rather, each effect is
described by a subspace of $\R^{27}$ of dimension 2 consisting of contrast
vectors, and the process of restriction yields a corresponding subspace of
$\R^9$.  We might denote the subspaces of $\R^{27}$ by $U_A, U_B, \ldots,
U_{AB^2C^2}$, and the corresponding subspaces of $\R^9$ by $\Uhat_A, \Uhat_B,
\ldots, \Uhat_{AB^2C^2}$.  The subspaces of $\R^{27}$ are mutually orthogonal.
The ``complete" aliasing $A = BC = ABC$ in the fraction means that $\Uhat_A =
\Uhat_{BC} = \Uhat_{ABC}$, while the fact that $A$ and $B$ are unaliased in the
fraction means that $\Uhat_A \perp \Uhat_B$.  Since this fraction is regular,
being equal (completely aliased) or orthogonal (unaliased) are the only
possibilities.

This is the way we will view aliasing in Section \ref{resolu}.  Two things
occur when we move to nonregular fractions.  One is the appearance of partial
aliasing.  The other is the lack of ``components of interaction" that we have
in regular $s$-level fractions when $s > 2$.

\section{Strength and resolution of fractional factorial
designs}\label{resolu}
Let $T$ be a finite set -- for us, a set of treatments.  An
observation on a treatment $s \in T$ is assumed to have a mean
$\mu(s)$, which we refer to as a \emph{cell mean} (when $T$ is a
Cartesian product, its elements are ``cells").  Contrasts in cell
means are expressions of the form
  \[ \sum_{s \in T} c(s) \mu(s) \]
where $\sum_{s \in T} c(s) = 0$.  We may refer to these functions
$c \in \R^T$ as \emph{contrast functions} or \emph{contrast
vectors}, or (by abuse of language) as \emph{contrasts}.

Any blocking (or partition) \sC\ of $T$ determines a subspace $U_{\ssC} \subset
\R^T$ of dimension $|\sC|-1$ consisting of the contrast functions that are
constant on the blocks of $\sC$. If $c \in U_{\ssC}$, then $\sum c(t) \mu(t)$
is a contrast between the blocks.  The association of a vector space $U_{\ssC}$
to each partition \sC\ was first formalized and studied by \citet{Tjur84}.

If \sD\ is another blocking of $T$, we define the \emph{join} of
\sC\ and \sD\ to be the partition
  \[\sC \vee \sD = \{C\cap D: \; C \in \sC, \; D \in \sD, \; C\cap D \neq
  \emptyset \}. \]

Let $\pi$ be the uniform probability measure on $T$:
  \[ \pi(A) = |A|/|T|. \]
We denote the independence of $A$ and $B$ by $A \pperp B$.  This
is simply the combinatorial condition
  \[ |A\cap B| |T| = |A| |B|. \]
We say that the set $A$ is independent of the partition \sC\ (written $A \pperp
\sC$) if $A \pperp C$ for every $C \in \sC$.  Similarly, the partitions \sC\
and \sD\ are independent ($\sC \pperp \sD$) if $C \pperp D$ for every $C \in
\sC$ and $D \in \sD$.  This condition is important because of the fact
\citep[Lemma 3]{Bed89} that
  \[ U_{\ssC} \perp  U_{\ssD} \;\; \mbox{iff} \;\;  \sC \pperp \sD.
  \]
Independence also gives us a convenient way to define the strength of
an orthogonal array (see  Lemma \ref{strength-lemma} below).

For the remainder of this section, let $T = \cart{A}{1}{k}$ be the
set of treatment combinations in an \cart{s}{1}{k} factorial,
where $A_i$ indexes the levels of factor $i$ and $s_i = |A_i|$.
Which main effect or interaction a contrast belongs to is
determined entirely by the coefficients $c(s)$.

As $r$ ranges over $A_i$, the sets
  \[\cart{A}{1}{i-1} \times \{r\} \times \cart{A}{i+1}{k} \]
form a blocking $\sA_i$ of $T$ consisting of $s_i$ blocks of equal
size. For $i<j$ the blocks of $\sA_i \vee \sA_j$ are sets of the
form
  \[\cart{A}{1}{i-1} \times \{r\} \times \cart{A}{i+1}{j-1} \times \{s\}
     \times \cart{A}{j+1}{k} \]
where $r \in A_i$ and $s \in A_j$.

In general, for any nonempty subset $I \subset \{1, \ldots, k\}$ the factors $i
\in I$ determine the blocking $\vee_{i \in I} \sA_i$ of $T$.  Its blocks are
formed by taking intersections of blocks, one from each $\sA_i$, $i \in I$, and
are subsets of $T$ of the form \cart{B}{1}{k}, where for fixed elements $r_i
\in A_i$ we have
  \begin{equation} \label{Bi-eq} B_i = \left \{ \begin{array}{rl}
   \{r_i\} &\mbox{if } i \in I, \\
     A_i &\mbox{if } i \not\in I.
     \end{array}\right. \end{equation}

We pause to record some simple observations that will be needed
below. Let $\sC_I$ denote $\veeA{I}$.
\begin{lemma}\label{lemma1}
\begin{enumerate}
  \item $\pi(B) = \displaystyle\frac{1}{\Pi_{i \in I} s_i}$ for every
  block $B \in \sC_I$.  \label{piB}
  \item $\sC_I \vee \sC_J = \sC_{I \cup J}$. \label{IcupJ}
  \item $\sC_I \pperp \sC_J \Leftrightarrow I \cap J = \emptyset$.
        \label{IcapJ}
\end{enumerate}
\end{lemma}
\begin{proof}
For $B \in \sC_I$, $|B|=\prod_{i=1}^k |B_i|=\prod_{i \not \in I }
s_i$. Thus $\pi(B)=\prod_{i \not \in I} s_i/|T|=\prod_{i \in I} s_i$.
This proves (\ref{piB}).

To prove (\ref{IcupJ}), let $B' \in \sC_I$ and $B'' \in \sC_J$.  Then
$B'=B'_1 \times \cdots \times B'_k$ and $B''=B''_1 \times \cdots
\times B''_k$ where $B'_i$ is of form (\ref{Bi-eq}) and  $B''_i$ is
of the same form with $I$ replaced by $J$ (and possibly different
elements $r_i$). We must show that either $B' \cap B''$ is also of
this form, $I \cup J$ replacing $I$, or $B' \cap B'' = \emptyset$.
But the first case occurs if $B'_i$ and $B''_i$ agree for all $i \in
I \cap J$ (trivially if $I \cap J = \emptyset$), while the second
occurs if they disagree.  Thus $\sC_I \vee \sC_J \subset \sC_{I \cup
J}$.

Conversely, if $B \in \sC_{I \cup J}$ then
$B$ is of form (\ref{Bi-eq}) with $I \cup J$ replacing $I$.  Using
the given values of $r_i, i \in I \cup J$, define
  \[ B'_i=\left\{\begin{array}{ll}
  \{ r_i \} & \mbox{if } i \in I,\\
  A_i & \mbox{if } i \not \in I.
  \end{array}\right.  \]
and
  \[ B''_i=\left\{\begin{array}{ll}
  \{ r_i \} & \mbox{if } i \in J,\\
  A_i & \mbox{if } i \not \in J
  \end{array}\right.  \]
and put $B'=B'_1 \times \cdots \times B'_k$ and $B''=B''_1 \times
\cdots \times B''_k$.  Note that $B'$ and $B''$ automatically agree on
$I \cap J$, and that $B' \in \sC_I$ and $B'' \in \sC_J$.  Then $B =B'
\cap B'' \in \sC_I \vee \sC_J $, and so $\sC_{I \cup J} \subset \sC_I
\vee \sC_J$, proving~(\ref{IcupJ}).

Finally, let $B' \in \sC_I$ and $B'' \in \sC_J$. If $I\cap J
=\emptyset$, then it is easy to see that $\pi(B' \cap
B'')=\pi(B')\pi(B'')$. If, however, $I \cap J \neq \emptyset$, then
either there exists $i \in I \cap J$ such that $r'_i \neq r''_i$, in
which case $\pi(B' \cap B'')=0$, or $\pi(B' \cap B'')= 1 / \prod_{i
\in I \cup J} s_i $. In either case, $\pi(B' \cap B'') \neq
\pi(B')\pi(B'')$ for $B' \in \sC_I$ and $B'' \in \sC_J$. This proves
(c).
\end{proof}

We now describe the contrasts belonging to main effects and to
various interactions in the factorial experiment.  First, the
contrasts between the blocks of $\sA_i$ define the main effect of
factor $i$. The set of such contrast functions is then
  \[  U_i = U_{\ssA_i}. \]
The contrast functions belonging to the $ij$-interaction are defined to be
those elements of $U_{\ssA_i \vee \ssA_j}$ that are orthogonal to both $U_i$
and $U_j$.  They form a subspace which we denote $U_{ij}$. In general, for
$\emptyset \neq I \subset \{1, \ldots, k\}$ we define the subspaces $U_I$
inductively as
  \[ U_I = \{c \in U_{\ssC}: c \perp U_J \; \mbox{for all} \; J
  \subsetneqq I \}, \]
where $\sC = \veeA{I}$ and  $U_{\emptyset}$ is the subspace of constant
functions.  For nonempty $I$, the subspace $U_I$ is the set of contrast
functions belonging to the interaction between the factors listed in the set
$I$.  This is a slightly modernized version of the definition given by
\citet{Bose47}.  We note that $U_{\ssC}$ has the orthogonal decomposition
\begin{equation} \label{decomp-eq}
  U_{\ssC} = \oplus_{J \subseteq I} U_J.
\end{equation}

Next, we consider what happens when we observe only those treatment
combinations in a subset, or \emph{simple fraction}, $S \subset T$.
\citet[page 129]{Rao47} referred to such subsets as \emph{arrays}.
His crucial discovery was the parameter known as \emph{strength}.

\begin{definition} \label{oa-def}  $S$ has \emph{strength $t\ge 1$}
if, for every $I=\{i_1, \ldots, i_t\}\subset \set{k}$, the projection
of $S$ onto the factors $i_1, \ldots, i_t$   consists of $\lambda_I$
copies of the full factorial $ A_{i_1}\times \cdots \times A_{i_t}$.
\end{definition}

Note that for a symmetric array $S$, the multiplicities $\lambda_I$
are all equal to a common value $\lambda$, the \emph{index} of the
array.

As is well known, it follows from the definition that if $S$ has strength $t$
then it also has strength $t'$ for all $t' < t$.  A convenient equivalent
definition of strength is the following \citep[Corollary 5.2]{bederrao}.

\begin{lemma} \label{strength-lemma}  A simple fraction $S$ has strength
$t$ iff for every $I \subset \set{k}$ of size $t$ we have
  \[ S \pperp \veeA{I}. \]
\end{lemma}

In order to define aliasing in a fraction $S \subset T$, we
restrict the contrast functions of the full factorial experiment to
the subset $S$.  Thus we let $\uhat$ be the restriction of $u$ to
$S$, and let
  \begin{equation}  \label{UhatI-eq} \Uhat_I = \{ \uhat: u \in U_I \}.
  \end{equation}
$\Uhat_I$ denotes the set of restrictions of all the functions in
$U_I$ to the fraction $S$. Since addition and scalar multiplication
are defined pointwise, $\Uhat_I$ is also a subspace (of $\R^S$).
The definition of aliasing that follows allows us to define resolution in
exactly the same way as \citet[page 319]{BoxHunt61} do in regular fractions.

\begin{definition} \label{res-def}  Let $S$ be a simple fraction.
$U_I$ and $U_J$ are \begin{itemize}
  \item[-] \emph{completely aliased in} $S$ if $\Uhat_I
= \Uhat_J$,
  \item[-] \emph{unaliased in }$S$ if $\Uhat_I \perp \Uhat_J$, and
  \item[-] \emph{partially aliased in} $S$ otherwise.
\end{itemize}

$S$ has \emph{resolution R} if, for each $p$, every $p$-factor effect
is unaliased with every effect having fewer than $R-p$
factors.\end{definition}

%

\noindent It is straightforward to see that
a fraction having resolution $R$ also has resolution $R'$ for all $R'
< R$.

We quote the following theorem and corollary from \citet[Theorem 3.4(a) and
Corollary 3.5]{bederrao2}. We include the brief proof of the corollary for
convenience.

\begin{theorem}\label{unaliased}
Let $S$ be a simple fraction of strength $t$. Let $I, J \subset \set{n}$ with
$|I \cup J| \le t$. If $I \neq J$, then $\Uhat_I \perp \Uhat_J$.
\end{theorem}

\begin{corollary}\label{coro}
If $S$ has strength $t$ then it has resolution $t+1$.
\end{corollary}

\begin{proof}
Suppose $S$ has strength $t$, and let $I$ and $J$ be subsets of
$\set{k}$ such that
  \[ |I| = p \quad \mbox{and} \quad |J| \le t-p. \label{res}  \]
By Theorem~\ref{unaliased}, $\Uhat_I \perp \Uhat_J$.  Thus no
interaction of $p$ factors is aliased with any interaction of at most
$t-p$ factors.  But this means that $S$ has resolution $t+1$.
\end{proof}

Corollary \ref{coro} implies that if $S$ has \emph{maximum} strength
$t$ then $S$ has resolution $R \geq t+1$.  We now show that $R$ cannot
exceed $t+1$.

\begin{theorem}\label{strength/resolution}
If a simple fraction $S$ has maximum strength $t$, then $S$ has maximum
resolution $t+1$.
\end{theorem}

\begin{proof} To show that $S$ does not have resolution $t+2$, we must
produce $I, J \subset \set{k}$ such that $|J| < t+2 - |I|$ but
$\hat{U}_I \not  \perp \hat{U}_J$.

Since $S$ does not have strength $t+1$, there exists a set $K
\subset \{1, \ldots, k\}$ such that $|K|=t+1$ and $S \npperp \sC_K$,
where $\sC_K = \bigvee_{i \in K} \sA_i$. That means there exists a
block $B \in \sC_K$ such that $S \npperp B$.

Now $|K| \ge 2$, so we may write $K=I \cup J$, where both $I$ and $J$
are nontrivial and $I \cap J =\emptyset$. Since $K=I \cup J$, we have
$\sC_K=\sC_I \vee \sC_J$ by Lemma \ref{lemma1}, so there exist $B' \in
\sC_I$ and $B'' \in \sC_J$ such that $B=B' \cap B''$.

Let $u = 1_{B'}-\pi(B')1$ and $v=1_{B''}-\pi(B'')1$. Then $u \in
U_{\ssC_I}$ and $v \in U_{\ssC_J}$.
Using equation (\ref{decomp-eq}) we have the orthogonal sums
  \[ u = \sum_{I' \subseteq I} u_{I'}, \;\; v =\sum_{J'
  \subseteq J} v_{J'}, \]
where $u_{I'} \in U_{I'}$ and $v_{J'} \in U_{J'}$.  Now if $I' \subset
I$ and $J' \subset J$, then $I' \neq J'$ (in fact they are disjoint);
moreover, if $I' \neq I$ or $J' \neq J$, then $|I' \cup J'| \leq t$,
and thus $(\hat{u}_{I'},\hat{v}_{J'})=0$ by Theorem \ref{unaliased}.
Hence $(\hat{u}, \hat{v})=(\hat{u}_I, \hat{v}_J)$. We will show that
$(\hat{u}, \hat{v}) \neq 0$. Then $(\hat{u}_I, \hat{v}_J) \neq 0$, and
thus $\hat{U}_I \not  \perp \hat{U}_J$. Now
  \[ \begin{array}{rcl} (\hat{u}, \hat{v})
  & = & \sum_{s \in S} u(s)v(s) \\
  & = & \sum_{s \in S} (1_{B'}(s)-\pi(B')1)(1_{B''}(s)-\pi(B'')1) \\
  & = & |B' \cap B'' \cap S|- \pi(B')|B'' \cap S|-\pi(B'')|B' \cap S|
        + |S|\pi(B')\pi(B'') \\
  & = & |T| \; [\pi(B' \cap B'' \cap S)-\pi(B')\pi(B'' \cap S)
        -\pi(B'')\pi(B' \cap S)+ \pi(S)\pi(B')\pi(B'') ].
  \end{array} \]
Since $S$ has strength $t$, it is independent of both $\sC_I$ and
$\sC_J$ (Lemma~\ref{strength-lemma}), and so we have $\pi(B' \cap
S)=\pi(B')\pi(S)$ and $\pi(B'' \cap S)=\pi(B'')\pi(S)$. Moreover,
since $I \cap J =\emptyset$, we have $\sC_I \pperp \sC_J$ by Lemma
\ref{lemma1} , and thus $\pi(B' \cap B'')=\pi(B')\pi(B'')$.
Therefore,
  \[  \begin{array}{rcl} (\hat{u}, \hat{v})
  & = & |T| \; [\pi(B' \cap B'' \cap S)-\pi(B')\pi(B'')\pi(S)] \\
  & = & |T| \; [\pi(B \cap S)-\pi(B')\pi(B'')\pi(S)] \\
  & = & |T| \; [\pi(B \cap S)-\pi(B' \cap B'')\pi(S)] \\
  & = &|T| \; [\pi(B \cap S) -\pi(B)\pi(S)] \\
  & \neq &0,
  \end{array}
  \]
since $S\npperp B$.
\end{proof}

In the following, the notation $OA(N,k,s,t)$ denotes a symmetric orthogonal
array of size $N$ (the number of ``runs"), $k$ factors, $s$ symbols and
strength $t$. 

%
%
%

\begin{example} \rm \label{4^4-ex} The solution set of the equation $x_1
+ x_2 + x_3 + 2x_4 \equiv 0 \pmod{4}$ forms a 1/4 fraction of a
$4^4$ factorial, and is an $OA(64,4,4,2)$.  It does not have
strength 3, as its projection on the first three factors is not a
complete $4^3$ factorial, but is rather the juxtaposition of 2
copies each of the fractions of a $4^3$ factorial given by $x_1 +
x_2 + x_3  \equiv 0 \pmod{4}$ and $x_1 + x_2 + x_3 \equiv 2
\pmod{4}$. Thus it has (maximum) resolution 3:  main effects are
unaliased with each other, but some two-factor interactions are
aliased with main effects.

We can even say a bit more about where the aliasing is occurring.
Let us call the factors $A, B, C$, and $D$. If we project the
fraction on any set of three factors that includes $D$, we indeed
get a complete $4^3$ factorial design. Thus those three main effects
and all their interactions will be unaliased in the fraction.
Aliasing between main effects and two-factor interactions is limited
to the three factors other than $D$.

One might be tempted to view this as a regular fraction with
defining contrast $ABCD^2$, based on the defining equation, and to
conclude that its resolution should be 4 since the ``wordlength" of
$ABCD^2$ is 4. However, the fraction is a solution set of an
equation over $\Z/4$, not $GF(4)$.
According to our definition, this fraction is not regular and the usual
wordlength algorithm need not apply -- indeed it doesn't.  We note that the
generalized wordlength pattern of this fraction, computed according to \citet{XuWu01},
is (0, 0, 1, 2).

\end{example}

\medskip
%
%
%
%

\begin{remark} \rm \label{strength-rem}

We have allowed fractions to have strength 1 (resolution 2). 
It is even possible to allow $t=0$ in our definition of strength
(take an empty join to be the trivial partition $\{T\}$). Then a
fraction has maximum strength 0 if it does not even have strength 1.
Similarly, all fractions vacuously have resolution 1, and with a bit
of elaboration one may show that all ensuing results extend to this
case. As there is no practical need for this, we have omitted it.

Roughly speaking, two contrasts are completely aliased if they are equal, and
unaliased if they are orthogonal.  Partial aliasing is something in between.
Using this idea, \citet[page 1074]{DengTang99} introduce a generalized
resolution for 2-level designs that equals Box-Hunter resolution plus a
fraction. This is extended to other designs by \citet[Section~5]{EKDD05}.

\end{remark}

\begin{remark} \rm  An alternate proof of Theorem~\ref{strength/resolution}
can be constructed using machinery introduced by \citet{ChengYe04}.
\end{remark} 

\section{Other approaches to resolution and aliasing}\label{approaches}

Several authors, such as \citet[page 152]{PWMJohn71}
and \citet[page 88]{RakHeFed}, define a fraction to have resolution $R$ if all
contrasts belonging to effects of order at most $[R/2] -1$ are estimable. (An
effect is of \emph{order} $k$ if it involves exactly $k$ factors.  The symbol
$[x]$ denotes the greatest integer not exceeding $x$.) This is accompanied by
the assumption that certain high-order effects are absent, an assumption not
present in Box and Hunter's original formulation (see \citet{bederrao2} for a
discussion of this issue). One consequence of their definition is that while a
design of strength $t$ has resolution $t+1$, it may, for example, have
resolution $R$ but maximum strength $R-2$ \citep[page 174]{RakHeFed}.
\citet[page 281]{HedayatSloaneStufken} add a further assumption to deal with
this problem.

\citet[page 18]{DeyMuk} define a fractional design to have resolution $(f,t)$,
$f \le t$, if contrasts belonging to effects of order at most $f$ are estimable
when effects of order greater than $t$ are absent.  They show that fractions of
strength $g$ are universally optimal as long as $f + t = g$, and then
\emph{define} the resolution of such a fraction to be $g+1$ \citep[Theorem
2.6.1 and Remark 2.6.2]{DeyMuk}.

Box-Hunter resolution distinguishes itself from these approaches in
a couple of ways:
\begin{itemize}

\item  It is a combinatorial property of the underlying fraction,
independent of any modeling assumptions (such as the absence of
high-order interactions).

\item  It is a measure of the amount of aliasing in a fraction
rather than of the estimability of terms in a model.

\end{itemize}

We are relying here on the concept of aliasing used in regular fractions
(Definition~\ref{res-def}).  We note, however, that aliasing is sometimes
defined in terms of bias, specifically as a measure of the biases caused by
misspecification of a model (cf. \citet[page 7]{BoxWil51} and \citet[page
95]{RakHeFed}). The relation between the two views of aliasing is discussed in
\citet{bederrao2}.




\section{Conclusion}
Consider the following simple orthogonal array, an $OA(18,3,3,2)$ that is the
juxtaposition of two regular $OA(9,3,3,2)$:
  $$O=\left[\begin{array}{cc}000111222&000111222\\
  012012012&120012210\\
  012120201&012012021\end{array}\right]$$
Here the runs are the 18 columns.  The components are the solutions to $x_1 +
x_2 + 2x_3 \equiv 0$ and $\equiv 1 \mod{3}$, and so the defining relation is $I
= ABC^2 = A^2B^2C$ for each component.  Thus the wordlength pattern for each
component is $(0,0,2)$, 
while the generalized wordlength pattern of the combined
array, computed according to \citet{XuWu01}, is $(0,0,1/2)$.

The question suggested by this example is just what information about aliasing
is encoded in the generalized wordlength pattern. We have seen that this
pattern does capture the resolution of a design in the sense of Box and Hunter.
What makes the wordlength pattern useful in the case of regular designs is the
underlying aliasing structure, which is directly controlled by the defining
words.  We do not yet seem to have a comparable theory for nonregular designs,
one that would explain an example such as the one above.  It would seem that
Theorem~\ref{strength/resolution} should be a natural part of such a theory.

The development leading up to our Theorem \ref{strength/resolution} depends
directly on the property that a design is simple (or equireplicate).  To extend
it to non-simple designs would require extending to such designs the concept of
aliasing in Definition~\ref{res-def}.  It is not clear at this time how to do
this.  Such an extension would be a larger goal of the theory.
\medskip

\textbf{Acknowledgments}.  We thank Jeb Willenbring and Dan Lutter for
assistance with computing.  We also thank the referees for correcting or clarifying references to the literature, and especially for drawing our attention to \citet{ChengYe04}.  Finally, we are grateful to Associate Editor Manohar L.~Aggarwal for his invaluable assistance.


\begin{thebibliography}{19}
\providecommand{\natexlab}[1]{#1}

\bibitem[{Beder(1989)}]{Bed89}
Beder, J.H., 1989. The problem of confounding in two-factor experiments.
  \emph{Communications in Statistics: Theory and Methods} A18, 2165--2188.
  Correction, A23(7), 1994.

\bibitem[{Beder(1998)}]{bederrao}
Beder, J.H., 1998. On {Rao's} inequalities for arrays of strength $d$.
  \emph{Utilitas Mathematica} 54, 85--109.

\bibitem[{Beder(2004)}]{bederrao2}
Beder, J.H., 2004. On the definition of effects in fractional factorial
  designs. \emph{Utilitas Mathematica} 66, 47--60.

\bibitem[{Bose(1947)}]{Bose47}
Bose, R.C., 1947. Mathematical theory of the symmetrical factorial design.
  \emph{Sankhy\={a}} 8, 107--166.

\bibitem[{Box and Hunter(1961)}]{BoxHunt61}
Box, G.E.P., Hunter, J.S., 1961. The $2^{k-p}$ fractional factorial designs.
  \emph{Technometrics} 3, 311--351; 449--458.

\bibitem[{Box and Wilson(1951)}]{BoxWil51}
Box, G.E.P., Wilson, K.B., 1951. On the experimental attainment of optimal
  conditions. \emph{Journal of the Royal Statistical Society, Series B} 13,
  1--45.

\bibitem[{Cheng and Ye(2004)}]{ChengYe04}
Cheng, S.W., Ye, K.Q., 2004. Geometric isomorphism and minimum aberration for
  factorial designs with quantitative factors. \emph{The Annals of Statistics}
  32, 2168--2185.

\bibitem[{Deng and Tang(1999)}]{DengTang99}
Deng, L.Y., Tang, B., 1999. Generalized resolution and minimum aberration
  criteria for {Plackett-Burman} and other nonregular factorial designs.
  \emph{Statistica Sinica} 9, 1071--1082.

\bibitem[{Dey and Mukerjee(1999)}]{DeyMuk}
Dey, A., Mukerjee, R., 1999. \emph{Fractional Factorial Plans}, Wiley, New
  York.

\bibitem[{Evangelaras et~al.(2005)Evangelaras, Koukouvinos, Dean, and
  Dingus}]{EKDD05}
Evangelaras, H., Koukouvinos, C., Dean, A.M., Dingus, C.A., 2005. Projection
  properties of certain three level orthogonal arrays. \emph{Metrika} 62,
  241--257.

\bibitem[{Fries and Hunter(1980)}]{FriesHunter80}
Fries, A., Hunter, W.G., 1980. Minimum aberration $2^{k-p}$ designs.
  \emph{Technometrics} 22, 601--608.

\bibitem[{Hedayat et~al.(1999)Hedayat, Sloane, and
  Stufken}]{HedayatSloaneStufken}
Hedayat, A.S., Sloane, N.J.A., Stufken, J., 1999. \emph{Orthogonal arrays:
  Theory and applications}, Springer Verlag, New York.

\bibitem[{John(1971)}]{PWMJohn71}
John, P.W.M., 1971. \emph{Statistical Design and Analysis of Experiments},
  Macmillan, New York.

\bibitem[{Ma and Fang(2001)}]{MaFang01}
Ma, C.X., Fang, K.T., 2001. A note on generalized aberration in factorial
  designs. \emph{Metrika} 53, 85--93.

\bibitem[{Raktoe et~al.(1981)Raktoe, Hedayat, and Federer}]{RakHeFed}
Raktoe, B.L., Hedayat, A., Federer, W.T., 1981. \emph{Factorial Designs}, John
  Wiley \& Sons, New York.

\bibitem[{Rao(1947)}]{Rao47}
Rao, C.R., 1947. Factorial experiments derivable from combinatorial
  arrangements of arrays. \emph{Journal of the Royal Statistical Society}
  Supplement, IX, 128--139.

\bibitem[{Tang and Deng(1999)}]{TangDeng99}
Tang, B., Deng, L.Y., 1999. Minimum {$G_2$}-aberration criteria for nonregular
  fractional factorial designs. \emph{The Annals of Statistics} 9, 1914--1926.

\bibitem[{Tjur(1984)}]{Tjur84}
Tjur, T., 1984. Analysis of variance models in orthogonal designs.
  \emph{International Statistical Review} 52, 33--81. With discussion.

\bibitem[{Xu and Wu(2001)}]{XuWu01}
Xu, H., Wu, C.F.J., 2001. Generalized minimum aberration for asymmetrical
  fractional factorial designs. \emph{The Annals of Statistics} 29, 1066--1077.

\end{thebibliography}

\bibliographystyle{jstp} 

\end{document}